\documentclass[aps,prc,twocolumn,superscriptaddress,nofootinbib]{revtex4-1} 
\usepackage{epsfig}
\usepackage{color}
\usepackage[latin1]{inputenc}
\usepackage{float,amsmath}
\usepackage{graphicx}

\newcommand{\ud}{\, \mathrm{d}}

\newcommand{\rt}{{\mathbf{r}_\perp}}

\newcommand{\xt}{{\mathbf{x}_\perp}}
\newcommand{\kt}{{\mathbf{k}_\perp}}
\newcommand{\yt}{{\mathbf{y}_\perp}}

\begin{document}

\title{Multiplicity distributions in p+p, p+A and A+A collisions from Yang-Mills dynamics}

\author{Bj\"orn Schenke}
\affiliation{Physics Department, Brookhaven National Laboratory, Upton, NY 11973, USA}
\author{Prithwish Tribedy}
\affiliation{Variable Energy Cyclotron Centre, 1/AF Bidhan Nagar, Kolkata 700064, India}
\author{Raju Venugopalan}
\affiliation{Physics Department, Brookhaven National Laboratory, Upton, NY 11973, USA}

\begin{abstract}
We compute transverse momentum and momentum integrated multiplicity distributions consistently in the IP-Glasma model for proton-proton and proton-lead  collisions at the LHC, in deuteron-gold collisions at RHIC, and in heavy ion collisions at both RHIC and LHC energies. Several sources of sub-nucleon scale contributions to the multiplicity distributions are identified. Our results, which are constrained by inclusive and diffractive deeply inelastic scattering data from HERA, are compared to measured distributions for a range of collision energies. These results are an essential first step in quantifying the relative role of initial and final state effects on multiparticle correlations in light and heavy ion collisions. 
\end{abstract}

\maketitle


\section{Introduction}

Computing bulk multiplicities and multiplicity distributions from first principles in QCD is extremely difficult. This is because multiplicities are dominated by transverse momenta where the physics may be intrinsically non-perturbative. If true, there is no small parameter to do systematic computations of such quantities. One is then forced to resort to simple models, at best weakly motivated by the underlying theory. From this perspective, the problem looks hopeless; it is especially chronic in heavy ion collisions because multiparticle production would on the surface appear more cumbersome than in elementary hadron-hadron collisions. 

The situation changed considerably with the advent of the Color Glass Condensate (CGC) effective field theory~\cite{Gelis:2010nm}. Gluon saturation suggests that a dynamical scale $Q_s\gg \Lambda_{\rm QCD}$ is generated in QCD at high energies, where $\Lambda_{\rm QCD}$ is  the underlying non-perturbative scale in the theory. If the coupling runs as a function of this dynamical scale, $\alpha_S(Q_s)\ll 1$, weak coupling methods could be used to compute quantities that were believed previously to be intractable. 

It is however difficult to completely avoid sensitivity to non-perturbative scales. This sensitivity is greatest for the most inclusive quantities, single inclusive multiplicities. In the CGC picture, the single inclusive multiplicity per unit rapidity in a high energy hadron-hadron collision can be 
expressed as $dN/d\eta = S_\perp Q_s^2/\alpha_S(Q_s)$. The numerator of the expression can be anticipated on purely dimensional grounds. The transverse overlap area of the collision $S_\perp$ is sensitive to non-perturbative dynamics.\footnote{If multiparticle production were purely non-perturbative, one would expect the multiplicity to go as $\Lambda_{\rm QCD}^2 S_\perp$, or a larger non-perturbative scale such as the string tension to replace $\Lambda_{\rm QCD}$. In some models, the string tension can be of the order of several GeV. However no simple understanding exists why such a scale should change with energy, as strongly favored by data.} While the multiplicative form of the $S_\perp$-dependence of the multiplicity suggests that non-perturbative scales may be ``factorized'' out, in practice they could have large residual effects. 

Nevertheless, because saturation models predict the energy and centrality dependence of $Q_s$, in principle there is a fair degree of predictive power in 
computations of single inclusive multiplicities\footnote{As we shall discuss in detail later, the dependence of $Q_s$ on energy and centrality cannot be claimed at present (optimistically) to be known to better than $10\%$ accuracy. This automatically implies minimally a $\sim 20\%$ uncertainty in multiplicities.}. In particular, an important quantity in understanding the event structure in high energy hadron-hadron collisions is the multiplicity distribution $P_n$. This quantity corresponds to the probability to produce $n$ charged hadrons in an event, either in full phase space or in restricted phase space domains. Charged particle multiplicity distributions in the central region of inelastic (non-single diffractive) $\bar{p}+p$ collisions at high energies were shown by the UA1 and UA5 collaborations to follow a negative binomial distribution \cite{Arnison:1982rm,Alner:1985zc}. Equivalently, multiplicity distributions in heavy ion collisions can be described by a superposition of negative binomial distributions (NBD) at different impact parameters.

A key observation is that negative binomial multiplicity distributions are naturally generated within the CGC framework \cite{Gelis:2009wh}. Negative binomial distributions (NBDs) are characterized by two parameters that describe the n-particle probability: these are ${\bar n}$, the mean multiplicity, and 
$k$, a parameter controlling the size of the fluctuations. For $k=1$, the distribution of $n$ particles is a highly correlated Bose-Einstein distribution; on the other hand, $k=\infty$ is a Poisson distribution corresponding to independent particle emission.  

In the CGC picture, highly occupied Glasma fields generated after the collisions are described as classical fields correlated in the transverse plane on correlation lengths of order $1/Q_s$, and are long range in rapidity. Remarkably, the multiplicity distribution generated by the decay of each of these ``Glasma flux tubes'' is close to a Bose-Einstein distribution~\cite{Dumitru:2008wn,Dusling:2009ar,Gelis:2009wh}. Since there are $S_\perp Q_s^2$ such flux tubes emitting gluons, the resulting distribution is a negative binomial distribution with $k\propto Q_s^2 S_\perp$. The Glasma flux tube picture has been explicitly confirmed in event-by-event numerical simulations of Yang-Mills equations \cite{Schenke:2012wb},  with the variance of the color charge fluctuations proportional to $Q_s^2$.

The Glasma framework of multiparticle production was compared (within in a $k_\perp$ factorization approximation to the Yang-Mills equations) to data from proton-proton collisions across a wide range of collision energies~\cite{Tribedy:2010ab} and with heavy ion collision data from RHIC~\cite{Tribedy:2011aa}. The results are in excellent agreement with data; however, because of the $k_\perp$ factorization employed, the results depend on and are sensitive to a fixed non-perturbative parameter $\zeta$. A more complete comparison to data in nucleus-nucleus collisions was performed in the CGC based IP-Glasma model~\cite{Schenke:2012wb,Schenke:2012hg,Gale:2012rq}, where the improvements included i) a more realistic implementation of nucleon fluctuations, and ii) solutions of Yang-Mills equations for the Glasma fields. The latter eliminated theoretical uncertainties in the $k_\perp$ factorization approximation--these can be significant for $k_\perp < Q_s$~\cite{Blaizot:2010kh}.

The fact that NBDs are intrinsic to the Glasma framework is advantageous for the IP-Glasma model relative to other models, such as the MC-Glauber model \cite{Miller:2007ri,Bozek:2013uha,Rybczynski:2013yba} or the MC-KLN model \cite{Dumitru:2012yr}. The latter models require the NBD fluctuations to be added by hand: In addition to $k$ being an additional tunable free parameter, the value of $k$ chosen is not dynamically reflected in the space-time structure of the fluctuations. 

This feature is included in the IP-Glasma model because the event-by-event sub-nucleon scale color charge fluctuations are constrained by the $Q_s$ extracted from HERA electron-proton deeply inelastic scattering (DIS) data. The extraction of $Q_s$ is performed within the framework of the IP-Sat dipole model~\cite{Bartels:2002cj,Kowalski:2003hm}. In this model, the dipole cross section is fit to HERA inclusive and exclusive data to obtain the dependence of $Q_s$ on Bjorken $x$ and impact parameter. Extending the dipole scattering framework to nuclei, one can extract ``lumpy'' color charge distributions of nuclei that are localized on transverse sizes  $\sim 1/Q_s$. As noted, these color charge configurations, via the Yang-Mills equations, generate the gluon fields that have the Glasma flux tube structure. The IP-Sat dipole model does not fully include the dynamics of multi-parton correlations, especially at very small values of $x$. The Glasma framework can however be systematically improved by including the multi-parton dynamics included in the Balitsky-JIMWLK evolution equations~\cite{Balitsky:1995ub,Jalilian-Marian:1997xn,Jalilian-Marian:1997jx,JalilianMarian:1998cb,Iancu:2000hn,Ferreiro:2001qy}.

Another important feature of the IP-Glasma+hydrodynamics framework is that it is one of the few ``event generators'' that includes the dynamics of the non-equilibrium stage\footnote{One such framework is the AMPT model~\cite{Zhang:1999bd,Pang:2013pma}, which has particle production via a string breaking mechanism, and re-scattering via Boltzmann transport processes.}. Presently, the framework utilizes only boost-invariant Yang-Mills evolution. However, recent developments~\cite{Berges:2012cj,Berges:2013eia,Epelbaum:2013waa,Gelis:2013rba,Berges:2013} suggest that extensions to solving 3+1-D Yang-Mills equations are possible, providing a qualitative improvement in treatment of the strongly correlated non-equilibrium Glasma stage.  

Both the implementation of the Balitsky-JIMWLK hierarchy and the 3+1-D Yang-Mills dynamics are left to future work. In this paper, we will focus on 
essential improvements to the IP-Glasma framework compared to previous implementations. 
While good agreement was obtained previously in comparison to multiplicity distributions in nucleus-nucleus collisions, the IP-Glasma model, as implemented in~\cite{Schenke:2012wb,Schenke:2012hg}, employed simplifying assumptions in constructing the  lumpy nuclear color charge distributions. One aim of this work is to eliminate these approximations to achieve a fully self-consistent treatment. 

Another goal is to extend the IP-Glasma model to address multiparticle production in smaller sized systems. Our interest in the latter is two-fold. Firstly, because the IP-Glasma model includes color fluctuations at the sub-nucleon level, it is important to test this picture for hadron-hadron and hadron-nucleus collisions. Secondly, since the IP-Glasma model was used successfully to provide the initial conditions for hydrodynamics in nucleus-nucleus collisions, it potentially provides a consistent framework to evaluate the relative contribution of initial and final state effects in nuclear collisions across a wide range in nucleon number, centrality and collision energy. Such a study is topical in light of recent results on two particle ``ridge'' correlations in proton/deuteron--nucleus collisions, that demonstrates great sensitivity of the results for both initial state~\cite{Dusling:2012iga,Dusling:2012cg,Dusling:2013oia}  and final state~\cite{Bzdak:2013zma,Bozek:2013uha,Bozek:2013ska} effects. A consistent description of $n$-particle distributions in different sized systems is essential for ``apples-to-apples'' comparisons of measurements.

The paper is organized as follows. In the next section, we will provide a brief introduction to the IP-Glasma model, emphasizing improvements to the model. In previous work, we took into account geometrical fluctuations in nucleon positions and sub-nucleon scale fluctuations in color charge density. (As noted, the latter generates a negative binomial distribution.)  An improvement over previous treatments is that we have incorporated high quality fits of the IP-Sat dipole model to the recent combined HERA data on inclusive and exclusive final states~\cite{Rezaeian:2012ji}. In addition to these sources of fluctuations, there are fluctuations in the Fock state configurations of projectile and target nucleons, for a given $Q_s$.  They correspond to fluctuations in gluon number in the nucleon wavefunctions \cite{Miettinen:1978jb}.\footnote{These fluctuations must be distinguished from fluctuations in the number of produced gluons in the final state, one source of which are the fluctuations in the charge density alone.}

In section \ref{sec:AA}, we present results on multiplicities in heavy ion collisions. We study in detail the dependence of the results 
on a) lattice spacing and lattice size, b) parameters that regulate the color fields in the infrared, c) different implementations of the running coupling constant, and d) the evolution time $\tau_0$ at which the gluon distributions are measured.\footnote{This time scale is also implicitly the time at which the matching to a hydrodynamic distribution will be performed. The hydrodynamic evolution will not be discussed in this work. Further detailed hydrodynamic studies are left to future work.} A generic feature of several of these systematic studies is the sensitivity of results to the number of participants $N_{\rm part}$. For smaller $N_{\rm part}$, the parameter dependence is greater, requiring more careful study. This points to the importance of cross-correlating studies of ``peripheral'' nucleus-nucleus collisions with those of p/d+A collision systems. In Sections \ref{sec:pp} and \ref{sec:pA}, we present results for proton-proton and proton-nucleus collisions. We demonstrate that additional sources of fluctuations, in excess of those described above, are important for describing rare events in 
these collisions. We also present results on single inclusive $p_T$ and rapidity distributions in proton-proton, proton/deuteron-nucleus collisions and compare these to available data. We summarize our results in Section \ref{sec:conc} and briefly discuss applications of our results to bulk phenomena in high energy hadron collisions.

\section{Fluctuations in the IP-Glasma}
Event-by-event fluctuations of the incoming nuclear wave functions have a large effect on observed particle multiplicity distributions.
The IP-Glasma model, described in detail in \cite{Schenke:2012wb,Schenke:2012hg}, includes several levels of fluctuations. They include fluctuation of nucleon positions in a nucleus and the sub-nucleon scale fluctuations of static large-$x$ color charges. These lead to fluctuations in the gluon fields that describe the dynamical small $x$ modes in the Color Glass Condensate (CGC) effective field theory~\cite{Gelis:2010nm}. The fluctuations in the gluon fields translate into fluctuations of global observables like the gluon multiplicity and transverse energy, as well as  fluctuations of the energy density in the plane transverse to the beam axis. The energy density fluctuations give rise to fluctuations of the charged hadron flow harmonics through the subsequent fluid-dynamical evolution of the system \cite{Gale:2012rq}. Measurements of multiplicity and flow distributions can therefore strongly constrain the non-equilibrium dynamics of the initial state. Conversely, better theoretical control of the initial conditions can help pin down transport coefficients in the quark-gluon-plasma (QGP). 

We shall briefly review key features of the IP-Glasma model that are relevant for the study of multiplicity fluctuations and discuss improvements 
over the model described in \cite{Schenke:2012wb,Schenke:2012hg}. As noted, an essential input is the dipole cross-section of the proton. The model we consider here is the IP-Sat saturation model \cite{Bartels:2002cj,Kowalski:2003hm}. The parameters of the model are fit to HERA data. In our earlier study, we used parameters from the fit in Ref.~\cite{Kowalski:2006hc}. However, high precision combined data from the H1 and ZEUS collaborations is now available. Excellent fits\footnote{Another dipole model, the b-CGC model, incorporating a different implementation of saturation physics, also has been shown recently to give very good agreement with the HERA data~\cite{Rezaeian:2013tka}. It will 
be interesting to see if there are observables that can cleanly distinguish between the different saturation scenarios.} of the combined inclusive data  are obtained in the IP-Sat model. With a further parameter governing the impact parameter dependence of the dipole cross-section, very good agreement is obtained with the HERA data on exclusive final states~\cite{Rezaeian:2012ji}. In this work we use the parameters found in \cite{Rezaeian:2012ji}.

The dipole cross-section for a nucleus for a given $x$ is constructed by taking the product of the S-matrices corresponding to the dipole cross-sections of overlapping nucleons at a given spatial location $\xt$. It can be expressed as~\cite{Kowalski:2007rw}
\begin{align}
&\frac{1}{2}\frac{\ud \sigma^{\textrm{A}}_{\textrm{dip}}}{\ud^2 \xt}(\rt,\xt,x)=\mathcal{N}_A(\rt,\xt,x)\notag\\
&~~~~=\left[1-e^{-\frac{\pi^2}{2N_{c}}\mathbf{r}_\perp^2\alpha_{s}(Q^{2}) xg(x,Q^{2})\sum_{i=1}^A T_p(\xt-\mathbf{x}_T^i)}\right]\,,
\label{eq:nuc-dipole}
\end{align}
where $T_p$ stands for the Gaussian thickness function for each of the $A$ nucleons in each nucleus. $\mathcal{N}_A$ is the scattering amplitude of the nucleus, $Q$ is the momentum scale related to the dipole size $\rt$, $Q^2=4/\mathbf{r}_\perp^2+Q_0^2$, with $Q_0$ fixed by the HERA inclusive data. 
The gluon distribution $xg(x,Q^{2})$ is parametrized at the initial scale $Q_0^2$ and then evolved up to the scale $Q^2$ using leading order DGLAP-evolution. The nuclear saturation scale $Q_s$ is the inverse value of $r=\sqrt{\mathbf{r}_\perp^2}$ for which $\mathcal{N} = 1 - e^{-1/2}$.
Because $Q_s$ is a function of both $\xt$ and $x$, and $x$ is given by the typical transverse momentum divided by $\sqrt{s}$, we need to self-consistently solve the equation $x = \lambda\,Q_s(\xt,x)/\sqrt{s}$ for every $\xt$, where we chose $\lambda=0.5$.\footnote{This method differs from that used in \cite{Schenke:2012wb,Schenke:2012hg}, where a constant $x=\langle p_\perp\rangle/\sqrt{s}$ was assumed. Note also that our results are insensitive to variations of $\lambda$ by factors of 2.}

The result of the outlined procedure is a lumpy distribution of $Q_s^2(\xt, x)$ in a nucleus. The expression in Eq.~(\ref{eq:nuc-dipole}) can also be computed in the McLerran-Venugopalan model (MV)~\cite{McLerran:1994ni,*McLerran:1994ka,*McLerran:1994vd}, thereby enabling one to relate the $Q_s^2$ extracted from data to the 
variance $g^2\mu^2$ of the color charge distribution of large x sources in the model. A quantitative relation between $Q_s^2$ and $g^2\mu^2$ was established in \cite{Lappi:2007ku}; while the coefficient relating the two can be computed numerically, it is sensitive to the parameters of the model.

It is important to note that the procedure followed here to extract $Q_s$ in the nucleus differs from that employed by us previously~\cite{Schenke:2012wb,Schenke:2012hg}. In the earlier work, we first determined $g^2\mu^2$ for individual nucleons which were then added. However this does not account for the different evolution speed in $x$ for heavy nuclei relative to that in the proton. In the IP-Sat model, the different speeds of evolution occur because the gluon distribution has a steeper $x$-dependence at the larger $Q_s^2$ scales reached in nuclei. Previously for simplicity we included this effect by hand. Determining $Q_s$ from the nuclear dipole cross section as described here, and using IP-Sat fits to deep inelastic scattering on nuclei, removes this freedom. 

The color charge fluctuations, as implemented in the MV model, are not the only source of initial state fluctuations in high energy QCD. 
The color charge fluctuations arise because a large number of gluons radiated in a fixed transverse area in a hadron can generate different representations of color charge. The distribution of color representations for large nuclei is well approximated by the MV model~\cite{Jeon:2004rk}. However a given $Q_s$ (which corresponds to the average $p_\perp$ kick experienced by a projectile probing the nucleus at high energies) can correspond to nuclear Fock state configurations with differing gluon number~\cite{Miettinen:1978jb}; this fluctuation is not accounted for in the MV model. For each such gluon number state with the same $Q_s$, one has a color charge distribution with a different variance. While we expect these fluctuations are not the dominant source of fluctuations in nucleus--nucleus collisions, they may be relevant in lighter systems. As we shall demonstrate in sections \ref{sec:pp} and \ref{sec:pA}, these fluctuations are important for a description of high multiplicity events in proton-proton and proton-nucleus collisions. 

A first principles understanding of the gluon number fluctuations in hadron wavefunctions remains challenging~\cite{Dumitru:2007ew}. To model these fluctuations, we shall relax the ``mean field'' assumption of strict linearity between $Q_s^2$ and $g^2\mu^2$ used in \cite{Schenke:2012wb,Schenke:2012hg}. This allows fluctuations of the gluon number, and hence color charge density, at a given $Q_s$. 
Specifically, we shall model this effect by a Gaussian distribution around the mean $g^2\mu^2$ and treat 
the width of the Gaussian as a free parameter. 

For a given $g^2\mu_{A(B)}^2(\xt,x)$ for nucleus A (B), the procedure follows that discussed previously \cite{Schenke:2012wb,Schenke:2012hg}.
To avoid numerical noise outside the interaction region, which can be significant in small systems like p+p and p+A collisions, we cut off exponential tails by setting $g^2\mu_{A(B)}^2(\xt,x)$ to be exactly zero wherever $T_p(\xt) < T_p^{\rm min}$. Unless otherwise noted,  $T_p^{\rm min}$ is chosen such that the maximal distance where $g^2\mu_{A(B)}^2(\xt,x)>0$ between the center of a nucleon at the edge of the interaction region and the edge is $r_{\rm max}\approx 1.2\,{\rm fm}$, approximately twice the gluonic radius of the proton $R_g\approx 0.6\,{\rm fm}$ \cite{Caldwell:2009ke}. This choice of $r_{\rm max}$ is large enough to avoid removing important contributions to the energy density and multiplicity.

One  samples $\rho_{A(B)}^a(\xt)$ in each event from a Gaussian distribution
\begin{equation}
\langle \rho_{A(B)}^a(\xt)\rho_{A(B)}^b(\yt)\rangle = g^2\mu_{A(B)}^2(x,\xt) \delta^{ab} \delta^{(2)}(\xt-\yt) \, ,
\end{equation}
and solves for the classical gluon fields in each nucleus using the Yang-Mills equations 
\begin{equation}\label{eq:YM1}
  [D_{\mu},F^{\mu\nu}] = J^\nu\,,
\end{equation}
where the color currents
\begin{equation}\label{eq:current}
  J^\nu_{A (B)} = \delta^{\nu \pm}\rho_{A (B)}(x^\mp,\xt)
\end{equation}
are generated by a nucleus A (B) moving along the $x^+$ ($x^-$) direction.
In (\ref{eq:current}), we chose a gauge where $A^\mp=0$,  with the result that temporal Wilson lines 
along the $x^+$ ($x^-$) axis become trivial unit matrices.

After solving Eq.\,(\ref{eq:YM1}) in Lorentz gauge $\partial_\mu A^\mu = 0$, where
\begin{equation}\label{eq:lor}
  A_{A(B)}^\pm = -\frac{\rho_{A (B)}(x^\mp,\xt)}{\boldsymbol{\nabla}_\perp^2+m^2}\,,
\end{equation}
the result can be transformed to light-cone gauge $A^+ (A^-) = 0$, where one finds 
 ~\cite{McLerran:1994ni,*McLerran:1994ka,*McLerran:1994vd,JalilianMarian:1996xn,Kovchegov:1996ty}   
\begin{align}\label{eq:sol}
  A^i_{A (B)}(\xt) &= \theta(x^-(x^+))\frac{i}{g}V_{A (B)}(\xt)\partial_i V^\dag_{A (B)}(\xt)\,,\\
  A^- (A^+) &= 0\,.\label{eq:sol2}
\end{align}
The infrared regulator $m$ in Eq.\,(\ref{eq:lor}) is of order $\Lambda_{\rm QCD}$ and crudely incorporates color confinement at the nucleon level. Because confinement is an intractable problem, it is not feasible to do better. The hope however is that physical observables are insensitive to $m$. 
We will discuss the dependence of our results on this mass term in section \ref{sec:AA}.

The initial condition for a high-energy nuclear collision at time $\tau=0$ is given by the solution of the CYM equations in Fock--Schwinger gauge 
$A^\tau=(x^+ A^- + x^- A^+)/\tau=0$. It has a simple expression in terms of the gauge fields of the colliding nuclei
\cite{Kovner:1995ja,Kovner:1995ts}:
\begin{align}
  A^i &= A^i_{(A)} + A^i_{(B)}\,,\label{eq:init1}\\
  A^\eta &= \frac{ig}{2}\left[A^i_{(A)},A^i_{(B)}\right]\,,\label{eq:init2}\\
  \partial_\tau A^i &= 0\,,\\
  \partial_\tau A^\eta &= 0
\end{align}

The numerical solution for these fields is discussed in \cite{Schenke:2012hg}, where we followed \cite{Krasnitz:1999wc,*Krasnitz:2000gz,Lappi:2003bi}.
The Glasma fields are then evolved in time $\tau$ by solving lattice discretized Hamilton's equations which are equivalent to the 
solution of Eq.\,(\ref{eq:YM1}) in the continuum limit. The numerical evaluation of the path-ordered exponential
\begin{equation}\label{eq:poe}
  V_{A(B)}(\xt)= \prod_{k=1}^{N_y} \exp \left(-ig\frac{\rho_k^{A(B)}(\xt)}{\boldsymbol{\nabla}_\perp^2+m^2}\right)
\end{equation}
involves the discretization of the longitudinal direction into $N_y$ steps. For the calculations used in this work, we use $N_y=100$ unless otherwise noted. The $N_y$ dependence of the results is shown in the appendix. 

To compute the gluon multiplicity per unit rapidity $dN_g/dy$ we fix transverse Coulomb gauge ($\partial_i A^i=0$, with $i$ summed over $1,2$). 
The lattice expression for $dN_g/dy$ is given by \cite{Krasnitz:2001qu,Lappi:2003bi}
\begin{align}\label{eq:N}
  \frac{dN_g}{dy} = \frac{2}{N^2} \int \frac{d^2k_T}{\tilde{k}_T} &\Big[\frac{g^2}{\tau} {\rm tr} \left(E_i(\kt) E_i(-\kt)\right)\nonumber\\
    & ~~ + \tau\, {\rm tr}
    \left(\pi(\kt)\pi(-\kt)\right)\Big]\,,
\end{align}
with $N$ being the number of lattice sites in one dimension. 
Here we assumed a free massless lattice dispersion relation for the interacting theory. 
This leads to the appearance of the square root of
\begin{equation}
  \tilde{k}_T^2 = 4 \left[\sin^2\frac{k_x}{2}+\sin^2\frac{k_y}{2}\right]\,,
\end{equation}
which is the effective lattice momentum squared.
The details of the solution of the Yang-Mills equations in this context are identical to those described in \cite{Schenke:2012hg}.

The solutions of the Yang-Mills equations are independent of the coupling constant. It enters only as a multiplicative factor in the final multiplicity.
Both terms in the square brackets in Eq.\,(\ref{eq:N}) are proportional to $1/g^2$ and by multiplying with $g^2/(4\pi \alpha_s(\tilde{\mu}))$ we introduce
running coupling effects employing a scale $\tilde{\mu}$.
Unless otherwise noted, we use $\tilde{\mu}=k_T/2$, where $k_T = |\kt|$ is the gluon momentum. We use the one-loop prescription for the running coupling
\begin{equation}\label{eq:running}
  \alpha_s(\tilde{\mu}) = \frac{4\pi}{\beta \ln\left[(\mu_0/\Lambda_{\rm QCD})^{2/c} + (\tilde{\mu}/\Lambda_{\rm QCD})^{2/c}\right]^c}\,.
\end{equation}
$\mu_0$ regulates the Landau pole. For the computations in the text, it is chosen to be $\mu_0=0.5\,{\rm GeV}$. The parameter $c$ controls the sharpness of this cutoff and is set to $c=0.2$. For $N_c=3$,  $\beta = 11-2 N_F/3$;  we set the number of flavors $N_F$ in this expression to 3. $\Lambda_{\rm QCD}$ is set to $0.2\,{\rm GeV}$. The dependence of the results on $\mu_0$, $c$ and the choice of scale $\tilde{\mu}$ is discussed later in the text and in the appendix. 

For p+p, d+Au, and p+Pb collision simulations, we use $N=400$ transverse lattice sites with lattice spacing $a=0.03\,{\rm fm}$, corresponding  
to a lattice of length $L=12\,{\rm fm}$. In A+A collisions, we use $N=600$ and $L=30\,{\rm fm}$ to accommodate the larger collision system.
We use an average ratio $Q_s/g^2\mu=0.65$ to convert the $Q_s$ from IP-Sat to a color charge density. We employ 
an infrared regulator $m=0.1\,{\rm GeV}$, $N_y=100$ and evolve for $\tau=0.4\,{\rm fm}$ before determining the multiplicity. Unless otherwise noted, the stated discretization and model parameters are the ones used in this study. The dependence on $N$, $L$, and the evolution time $\tau$ is examined in the appendix. 

 We note that while there are a large number of lattice parameters and model parameters in this study, most of these only significantly affect the overall normalization. The parameters that have an effect on the energy and centrality dependence of our results, namely $m$, $\tilde{\mu}$, and the ratio $Q_s/g^2\mu$, are discussed later in the text.

\section{Multiplicities in nucleus-nucleus collisions}\label{sec:AA}
Because we have introduced several improvements over the calculations presented in \cite{Schenke:2012wb,Schenke:2012hg} we first revisit the calculation of the multiplicity distributions in heavy ion collisions.
Essential differences to previous work include i) the cutoff of the color charge density at $r_{\rm max}$ away from the center of a nucleon at the edge, ii) the running coupling with $k_T$, and iii) the inclusion of the iteratively solved relation $x=\lambda\,Q_s(\xt,x)/\sqrt{s}$.  Pursuant to the previous discussion on fluctuations in the ratio $g^2\mu /Q_s$, these are not included for nucleus-nucleus collisions.  Technically this is because it should be implemented locally for a nucleus, which has not been done. However, in the dense environment of nuclear collisions, we anticipate these fluctuations have only a small effect on the multiplicity distributions.

The results in the the IP-Glasma model are for gluons, and are computed for the momentum space rapidity $y$.  To translate these results into 
results for the pseudo-rapidity $\eta$ requires a Jacobian factor. Here we follow the parametrization of \cite{Albacete:2012xq}, 
\begin{equation}
  \frac{dN_{\rm ch}}{d\eta} = \frac{\cosh \eta }{\sqrt{\cosh^2\eta+m_{\rm eff}^2/P^2}} \frac{dN_{\rm ch}}{dy}\,,
\end{equation}
with $m_{\rm eff}=0.35\,{\rm GeV}$ and $P = 0.13\,{\rm GeV}+0.32\,{\rm GeV}(\sqrt{s}/(1\,{\rm TeV}))^{0.115}$.
This transformation is of course approximate and introduces some uncertainty, because we are converting massless gluons to various massive hadrons represented by an effective mass $m_{\rm eff}$ and effective momentum $P$. 

\begin{figure}[tb]
   \begin{center}
     \includegraphics[width=8cm]{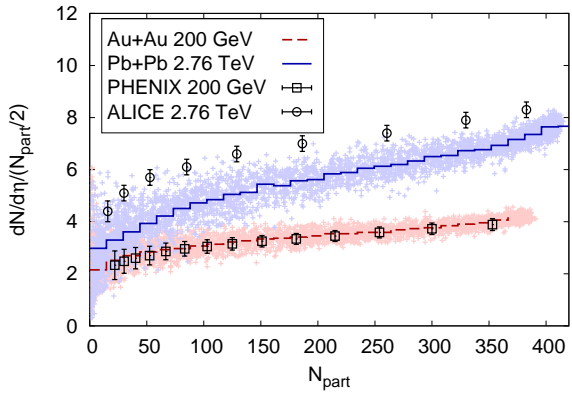}
     \caption{(Color online) Charged particle multiplicity divided by $N_{\rm part}/2$ as a function of $N_{\rm part}$
       compared to experimental data from the PHENIX \cite{Adler:2004zn} and the ALICE \cite{Aamodt:2010cz} collaborations.
       The bands are a collection of the multiplicities for individual events, 
       with the solid lines representing the average multiplicity.}
     \label{fig:dNdy-kt}
   \end{center}
\end{figure}

Fig.\,\ref{fig:dNdy-kt} shows the new results for the charged particle multiplicity as a function of $N_{\rm part}$ in Au+Au collisions at RHIC and Pb+Pb collisions at the LHC. 
We underestimate the Pb+Pb data at the higher $\sqrt{s}$ after fixing the normalization for Au+Au at $\sqrt{s}=200\,{\rm GeV}$. At low $N_{\rm  part}$ the result is about $30\%$ too low, at large $N_{\rm part}$ only $15\%$. Note however that additional entropy can be generated at the end of the Glasma stage. In particular, increased entropy production at higher energy (larger effective $\eta/s$) and in more peripheral events could provide a natural explanation for this difference. In \cite{Gale:2012rq}, the IP-Glasma Yang-Mills dynamics was matched event-by-event  to the \textsc{music} relativistic hydrodynamical model~\cite{Schenke:2010nt,Schenke:2010rr,Gale:2013da}. We will test in the future whether entropy production in \textsc{music}, when combined with these IP-Glasma results, can account for the additional entropy that appears to be required for agreement with the LHC data. 

\begin{figure}[htb]
   \begin{center}
     \includegraphics[width=8cm]{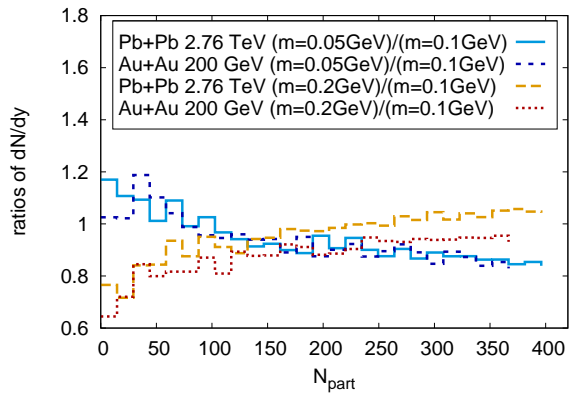}
     \caption{(Color online) Dependence of multiplicities on the infrared cutoff $m$. Results are most sensitive at small $N_{\rm part}$.
       Here $r_{\rm max}=0.97\,{\rm fm}$, $N_y=10$, and the coupling runs with $\tilde{\mu}=0.5\,k_T$.}
     \label{fig:dNdy-m}
   \end{center}
\end{figure}

\begin{figure}[htb]
   \begin{center}
     \includegraphics[width=8cm]{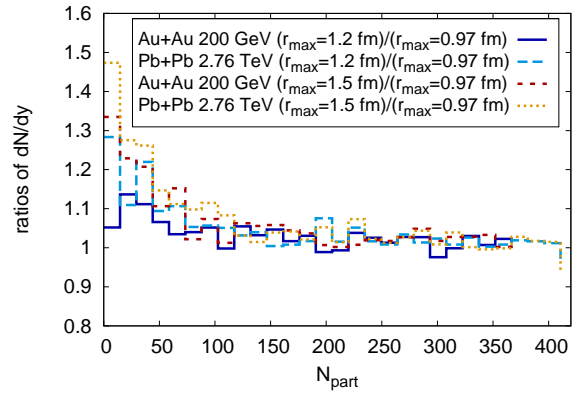}
     \caption{(Color online) Dependence on the cutoff on the color charge density at distance $r_{\rm max}$ from the nucleon center. Here, $m=0.1\,{\rm GeV}$, $N_y=10$, and the coupling runs with $\tilde{\mu}=0.5\,k_T$.}
     \label{fig:dNdy-Tp}
   \end{center}
\end{figure}

We will now investigate the dependence of the results presented in Fig.\,\ref{fig:dNdy-kt} on different parameters.
We shall first show how the multiplicity as a function of $N_{\rm part}$ is affected by the infrared cutoff $m$ on the gauge fields in the incoming nuclei in Eq.\,(\ref{eq:lor}). The result is shown in Fig.\,\ref{fig:dNdy-m} (We chose $N_y=10$ in this calculation.) We plot the multiplicity for $m=0.05\,{\rm GeV}$ and $m=0.2\,{\rm GeV}$ relative to a baseline value of $m=0.1$ GeV. We find that a larger mass term leads to a steeper slope of $dN/d\eta$ as a function of $N_{\rm part}$. While a mass of $100\,{\rm MeV}$ describes the RHIC data best (see Fig.\,\ref{fig:dNdy-kt}), a smaller mass term is preferred for better agreement with the LHC data. 

We next discuss the effect of the minimum value $T_p^{\rm min}$ at which we cut off the color charge density distribution $T_p$ to avoid numerical noise at large distances as discussed in the previous section.
In the IP-Sat model, the thickness function is parametrized as
\begin{equation}
  T_p(x,y) = \frac{1}{2\pi B_G} e^{-(x^2+y^2)/(2B_G)}\,,
\end{equation}
where $B_G = 4\,{\rm GeV}^{-2}$ gives good agreement with HERA exclusive vector meson and diffractive data \cite{Rezaeian:2012ji,Marquet:2008ha}.
For clarity, we characterize the cutoff $T_p^{\rm min}$ by the distance from the center of a nucleon, at which it cuts off the tail of the distribution.
This distance is $r_{\rm max}=\sqrt{-2 B_G \ln(2 \pi B_G T_p^{\rm min})}$. Note that in nuclei we only cut the distribution when we are at the edge of the interaction region. This is achieved by first adding all nucleons' $T_p$ and then determining whether the total $T_p<T_p^{\rm min}$.

The result is shown in Fig.\,\ref{fig:dNdy-Tp}. At large $N_{\rm part}$ we are completely insensitive to the cutoff, as we would expect. However we find a difference at low $N_{\rm part}$, with the multiplicity increasing with larger $r_{\rm max}$. This is in line with the results shown in Fig.\,\ref{fig:dNdy-m}, since larger $m$ leads to a smaller interaction region.
In previous calculations of flow in heavy ion collisions \cite{Gale:2012rq} we used $r_{\rm max}=\infty$. Because we did not study events with $N_{\rm part}<100$, the results obtained in \cite{Gale:2012rq} are completely unaffected by the introduction of a finite $r_{\rm max}$. These studies however tell us that how one treats intrinsically infrared physics can impact the conclusions one draws in peripheral nucleus-nucleus collisions. 

\begin{figure}[htb]
   \begin{center}
     \includegraphics[width=8cm]{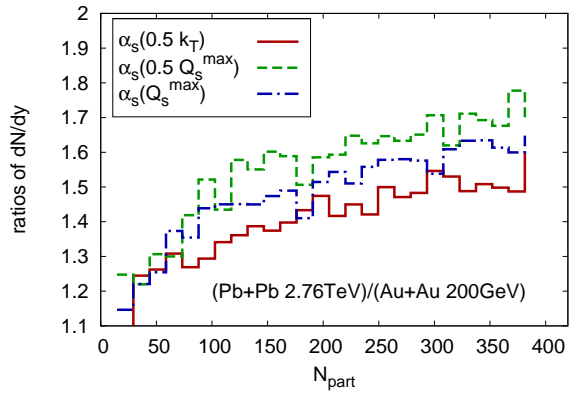}
     \caption{(Color online) Dependence on the scale used in the running coupling. Shown are ratios of the multi\-plicity of Pb+Pb collisions at $2.76\,{\rm TeV}$
       to Au+Au collisions at $200\,{\rm GeV}$. $m=0.1\,{\rm GeV}$ and $r_{\rm max}=1.2\,{\rm fm}$. $N_y=10$.}
     \label{fig:dNdy-running}
   \end{center}
\end{figure}

In Fig.\,\ref{fig:dNdy-running} we present the effect of the choice of scale $\tilde{\mu}$ in the running coupling. We remind the reader that the inverse coupling $\alpha_s^{-1}(\tilde{\mu})$ enters as an overall factor in the final multiplicity.
To emphasize the effect of the scale $\tilde{\mu}$ on the energy dependence of the multiplicity, we show the ratio of results for Pb+Pb collisions at $2.76\,{\rm TeV}$ to Au+Au collisions at $200\,{\rm GeV}$.  We compare $\tilde{\mu} = 0.5\, k_T$, with $k_T$ being the transverse momentum of the produced gluon, to $\tilde{\mu} = 0.5\, Q_s^{\rm max}$ and $\tilde{\mu}= Q_s^{\rm max}$, where $Q_s^{\rm max}$ is the larger of the two nuclei's $Q_s$ at every transverse position $\xt$. We find the strongest energy and centrality dependence for $\tilde{\mu} = 0.5\,Q_s^{\rm max}$, the weakest for $\tilde{\mu}\propto k_T$, with the relative difference being $\sim 15\%$ at larger centralities and diminishing at the lowest centralities where the running coupling is frozen, as suggested by the functional form in Eq.\,(\ref{eq:running}). 

\begin{figure}[htb]
   \begin{center}
     \includegraphics[width=8cm]{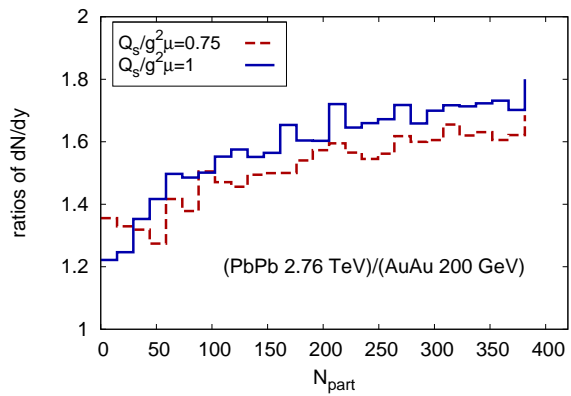}
     \caption{(Color online) Dependence on the ratio of $Q_s$ to $g^2\mu$, which mainly affects the overall normalization. We show the ratio of the result for Pb+Pb collisions at $2.76\,{\rm TeV}$ to Au+Au collisions at $200\,{\rm GeV}$ to demonstrate the effect of different values on the energy dependence. $m=0.1\,{\rm GeV}$ and $r_{\rm max}=1.2\,{\rm fm}$. Running with $\tilde{\mu}=0.5\,k_T$, $N_y=50$.}
     \label{fig:dNdy-gmu}
   \end{center}
\end{figure}

A similar plot showing the dependence of the energy evolution on the ratio of $Q_s$ and $g^2\mu$ is shown in Fig.\,\ref{fig:dNdy-gmu}. A larger value of  $Q_s/g^2\mu$ leads to a slightly stronger energy dependence. Compared to the large effect on the normalization, which is approximately a factor of two larger for the smaller value $Q_s/g^2\mu=0.75$, this effect on the energy evolution is small.

Further studies of parameter dependencies are presented in Appendix \ref{sec:app1}.
The only parameters  in our study that have an effect on the energy dependence of our results are the ones we discussed above, namely $m$, $\tilde{\mu}$, and the ratio $Q_s/g^2\mu$. Within the parameter ranges studied, the energy dependence of multiplicities varies by maximally $\sim 15\%$. 
These three parameters also have an effect on the $N_{\rm part}$ dependence, especially at $N_{\rm part} \lesssim 100$. However, the variation is at most $20\%$.\footnote{At very low $N_{\rm part}$ a larger effect is found when varying $r_{\rm max}$. However, this is attributable to numerical noise at large distances from the interaction region.}

 In the following sections, we will apply the IP-Glasma model to study multiplicity distributions in proton-proton and proton/deuteron-nucleus collisions. One objective of our study is to determine whether a consistent parameter set can be found that constrains the dynamics of the smaller size systems and those of peripheral nucleus-nucleus collisions.

\section{Multiplicities in proton+proton collisions}\label{sec:pp}

\begin{figure*}[tb]
   \begin{center}
     \includegraphics[width=12cm]{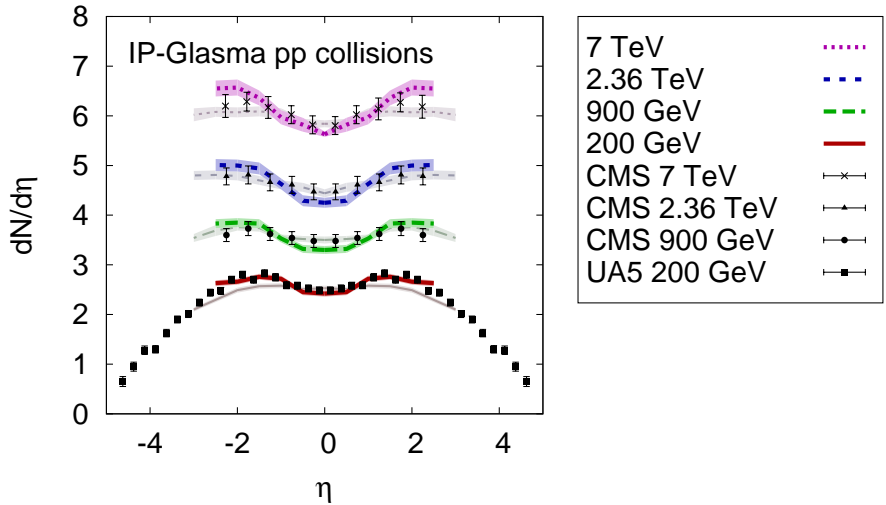}
     \caption{(Color online) Charged particle multiplicity as a function of pseudo-rapidity
       compared to experimental data from the UA5 \cite{Ansorge:1988kn} and the CMS collaboration \cite{Khachatryan:2010us}. Thick (colored) lines
     correspond to the best parameter set for AA collisions. Thin (gray) lines use $m_{\rm eff}=200\,{\rm GeV}$ which makes the dip around $\eta=0$ less prominent, and $N_y=10$, $\tau=0.5\,{\rm fm}$, and $Q_s/g^2\mu=0.75$.}
     \label{fig:dNdeta-pp}
   \end{center}
\end{figure*}

The overlap function of a proton+proton collision at impact parameter $b$ is a convolution of the corresponding thickness functions
\begin{align}
  T_{pp}(b) = \int dx \,dy\, T_p^A(x+b/2,y) T_p^B(x-b/2,y)\,.
\end{align}
With this quantity in hand, we can define the probability density for an inelastic parton-parton interaction as a function of the impact parameter. This is parametrized as 
\begin{equation}\label{eq:dPdb}
  \frac{d^2P}{d^2b}(b) = \frac{1-e^{-\sigma_{gg} N_g^2 T_{pp}(b)}}{\int d^2b\,\left( 1-e^{-\sigma_{gg} N_g^2 T_{pp}(b)}\right)}\,,
\end{equation}
with an effective parton number $N_g$ and effective parton-parton cross section $\sigma_{gg}$ \cite{d'Enterria:2010hd}.
The denominator of Eq.(\ref{eq:dPdb}) is the inelastic proton+proton cross section $\sigma_{pp}^{\rm inel}$, and we 
fix the value of $\sigma_{gg} N_g^2$ to reproduce its experimentally determined value $\sigma_{pp}^{\rm inel} = 68\,{\rm mb}$ for $\sqrt{s}=7\,{\rm TeV}$  
($\sigma_{pp}^{\rm inel} = 42\,{\rm mb}$ for $\sqrt{s}=0.2\,{\rm TeV}$) \cite{Chatrchyan:2012nj}.
In practice, we sample $b$ from a uniform distribution between $b_{\rm min}=0\,{\rm fm}$ and $b_{\rm max}=4\,{\rm fm}$ and weight each event with the factor $b\left(1-e^{-\sigma_{gg} N_g^2 T_{pp}(b)}\right)$.

The expression in Eq.~(\ref{eq:dPdb}) describes the likelihood of an inelastic proton-proton collision. Given such a collision, we next follow the procedure described earlier to compute the configuration of gauge fields created in the collision, and from these gauge fields, the gluon multiplicity using Eq.\,(\ref{eq:N}). We first show our results for the single inclusive rapidity and $p_T$ distributions. 

Before we do so, note that there is a constant normalization factor accounting for the difference in gluon vs. charged hadron number. It further absorbs normalization uncertainties coming from the choice of the ratio of $Q_s$ to $g^2\mu$, the infrared regulator $m$, the value of $N_y$ and $r_{\rm max}$, the choice of the running coupling scale, and dependencies on the lattice spacing $a$ (see Appendix \ref{sec:app1} for details). In principle, this factor should be energy independent. However we find that the needed normalization constant depends logarithmically on $\sqrt{s}$.
For the parameters employed, we find that this normalization factor $\mathcal{N}$ between the charged particle and the gluon multiplicity, $dN_{\rm ch}/d\eta = \mathcal{N} dN_{\rm g}/d\eta$, is $\mathcal{N} \approx 0.165 \ln(\sqrt{s}/1\,{\rm GeV})$. The need for such energy dependent normalization can possibly be understood from the fact that the IP-Sat model does not account for Gribov-diffusion - the growth of the nucleon with energy \cite{Gribov:1973jg}.
The growth of the transverse size from this diffusion is in fact expected to be proportional to $\ln s$.\footnote{This feature may be better accounted for in the b-CGC implementation of saturation physics \cite{Iancu:2003ge,Kowalski:2006hc,Watt:2007nr}. However, the b-CGC model is difficult to implement for nuclei.}

In the calculations presented, the coupling runs with the produced gluon's $k_T$. Using the maximal $Q_s(\xt)$ as a scale for the running coupling leads to a stronger energy dependence as shown in previous calculations in A+A collisions. (See Section \ref{sec:AA} and Ref.\,\cite{Schenke:2012hg}.) However our studies for asymmetric hadron collisions (discussed in Section \ref{sec:pA}) indicate that the choice of the running coupling scale can lead to results at variance with the measured $\eta$ dependence in these collisions. Thus while the energy dependence of the multiplicities prefers the choice of running with $Q_s$, the rapidity distributions in proton-nucleus collisions strongly favor the choice of running with $k_\perp$. This suggests a strong dependence of the results on the choice of scale, and on how the running coupling is frozen out at soft momenta.  While a particular choice may work for one quantity, it can fail for another. 
The sensitivity of the pseudo-rapidity dependence of the multiplicity on the choice of scale in the running coupling is troublesome on the surface.
However strictly speaking, running coupling effects enter at higher order in multiplicity computations. So a fully consistent treatment should include other next-to-leading-order corrections in addition to those absorbed in the running of the coupling. We would then expect the resulting expressions to be less sensitive to the choice of scale. For work in this direction, see \cite{Horowitz:2010yg}.

Our results for $dN_{\rm ch}/d\eta$ are shown in Fig.\,\ref{fig:dNdeta-pp}, with statistical errors indicated by bands. With the above stated caveats, 
we see that the IP-Glasma model gives a good description of the energy and rapidity distribution. 
Thick (colored) lines are for the parameters that gave the best description in A+A collisions. Thin (gray) lines are for a different set, using  
$m_{\rm eff}=200\,{\rm GeV}$, $N_y=10$, $\tau=0.5\,{\rm fm}$, and $Q_s/g^2\mu=0.75$. The main difference is caused by the smaller $m_{\rm eff}$ in the Jacobian, which makes the dip around $\eta=0$ weaker.
We are able to determine the multiplicity as a function of rapidity only on average because the result in a single event is strictly boost-invariant.
However, varying the rapidity $y$ in $x=(Q_s(\xt,x)/\sqrt{s})\exp(\pm y)$ will vary the magnitude of $Q_s$ in both protons in opposite ways--this feature of the model leads to the observed rapidity dependence of the multiplicity.

To check whether we can also reproduce the experimentally determined charged particle multiplicities as a function of transverse momentum, 
we compute the charged hadron distribution from the gluon distribution using the next-to-leading order (NLO) KKP \cite{Kniehl:2000fe} 
fragmentation functions:
\begin{equation}\label{eq:frag}
  \frac{dN^{h}}{dy d^2p_T} = \int_{0.05}^{1} \frac{dz}{z^2} D_g^{h}\left(z=\frac{p_T}{k_T},Q=k_T\right)\frac{dN^g}{dyd^2k_T}\,,
\end{equation}
where $D_g^h(z,Q)$ is the probability to produce a charged hadron with momentum $p_T=z k_T$ from a gluon with momentum $k_T$ at the scale $Q$.
We have restricted the integral to $z\geq 0.05$ so that the fragmentation function parametrizations are not used too far outside the $x$
range selected for the fits \cite{Kniehl:2000fe}.

\begin{figure}[htb]
   \begin{center}
     \includegraphics[width=8cm]{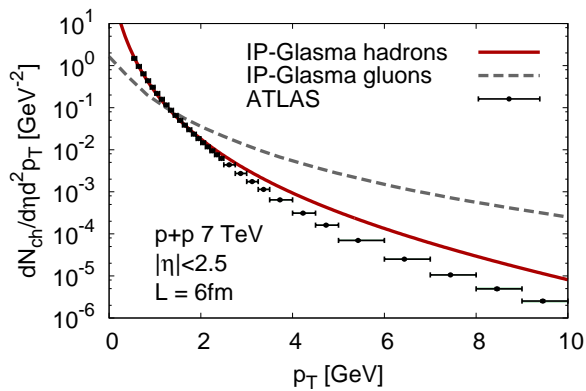}
     \caption{(Color online) Charged particle multiplicity obtained using Eq.\,(\ref{eq:frag}) as a function of transverse momentum
       compared to experimental data from the ATLAS collaboration \cite{Aad:2010ac}.}
     \label{fig:Npt}
   \end{center}
\end{figure}

The result is presented in Fig.\,\ref{fig:Npt}. Here we employ a lattice spacing of $a=0.015\,{\rm fm}$ to increase the momentum range on the lattice to higher momenta. We show both the gluon distribution and the charged hadron distribution after fragmentation.
The agreement with experimental data from the ATLAS collaboration \cite{Aad:2010ac} is very good at low momenta. At higher transverse momentum ($\gtrsim 3$ GeV), we see a similar overestimation of the experimental data as found in calculations using the McLerran-Venugopalan (MV) model \cite{Albacete:2010ad}. Since we have not attempted to introduce an anomalous dimension, or similar modification \cite{Albacete:2009fh,Lappi:2013zma}, or higher order $\rho^a$-correlators \cite{Dumitru:2011ax}, we anticipated the somewhat harder spectra at large momentum.

Furthermore, the quark contribution is ignored completely. It is well known that quark-gluon scattering can provide a significant contribution to the multiplicity as $x\rightarrow 0.01$ and above~\cite{Sassot:2010bh}. Note that the normalization factor is now somewhat different from the previous result for $dN/d\eta$, because the conversion from gluons to hadrons is taken care of by the fragmentation function.
Given these limitations, the overall agreement with the data is in fact surprisingly good. The shape of the spectrum in the range $0.5\,{\rm GeV}<p_T<2\,{\rm GeV}$ is well reproduced.

\begin{figure}[htb]
   \begin{center}
     \includegraphics[width=8.5cm]{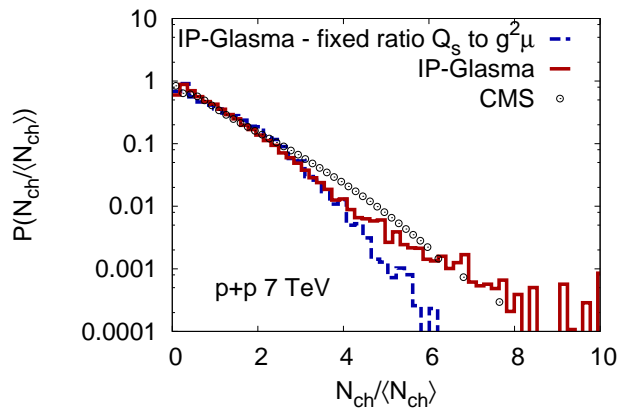}
     \caption{(Color online)  Minimum-bias charged particle multiplicity distribution scaled by the mean multiplicity compared to 
       experimental data from the CMS collaboration \cite{Khachatryan:2010nk}. We show the distribution with (solid) and without (dashed) smearing of 
the relation between $Q_s$ and the color charge density $g^2\mu$.}
     \label{fig:dNdy-pp}
   \end{center}
\end{figure}

Finally, we show the multiplicity distribution of p+p collisions at $7\,{\rm TeV}$ scaled by the mean multiplicity in Fig.\,\ref{fig:dNdy-pp}.
One clearly sees that when fixing the ratio of $Q_s$ to the color charge density $g^2\mu$, the distribution is too narrow, missing fluctuations in the tail of the distribution. As we discussed previously, there are sources of fluctuations in QCD that go beyond those included in our framework -- for a recent discussion, see \cite{Dumitru:2012tw} and references therein. If we allow $g^2\mu$ to fluctuate around its mean value with a Gaussian distribution whose width is $9\%$ of that mean value, the result is closer to the experimental data. As discussed above, the Gaussian distribution is merely an ansatz chosen for simplicity. The discrepancy between our result and the experimental data suggests that the exact form of these fluctuations is non-Gaussian.

The ridge correlation (a two particle correlation collimated at relative azimuthal separation $\Delta\Phi\approx 0$, and long range in their relative 
rapidity separation) observed in proton-proton collisions~\cite{Khachatryan:2010gv}  occurs in rare high multiplicity events. Their description in the CGC framework requires one crank up the saturation scale well beyond what the impact parameter dependent IP-Sat fits to the HERA data would give~\cite{Dusling:2012iga,Dusling:2012cg,Dusling:2013oia}. Our study allows for future computations where two-particle correlations, in a given centrality selection, are computed consistently for the appropriate corresponding saturation scales. 

\section{Multiplicities in proton+nucleus collisions}\label{sec:pA}
Proton-lead collisions at center of mass energies of $\sqrt{s}=5020\,{\rm GeV}$ have recently been performed at the LHC. d+A collisions at $\sqrt{s}=200\,{\rm GeV}$ were performed previously at RHIC.
We follow the same procedure for computing the multiplicity in p+A as in p+p collisions, using the weight in Eq.\,(\ref{eq:dPdb}). We then 
compute the configuration of gauge fields created in the collision and from those the gluon multiplicity according to Eq.\,(\ref{eq:N}).

\begin{figure}[htb]
   \begin{center}
     \includegraphics[width=8.5cm]{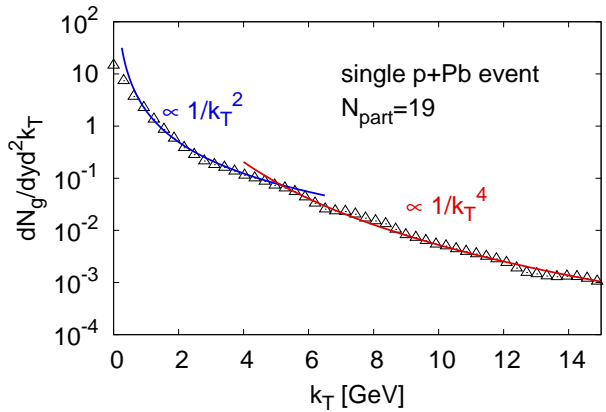}
     \caption{(Color online) Gluon transverse momentum distribution in a single p+Pb event at $\sqrt{s}=5020\,{\rm GeV}$ with $N_{\rm part}=19$.
       The solid line indicates the functional behavior $\sim 1/k_T^2$, the dashed line $\sim 1/k_T^4$.}
     \label{fig:kT}
   \end{center}
\end{figure}

First, we show the unintegrated distribution $dN_g/dyd^2k_T$ in transverse Coulomb gauge for a single event in Fig.\,\ref{fig:kT}.
We see that at large $k_T \gtrsim 5\,{\rm GeV}$ we recover the standard perturbative behavior $\sim 1/k_T^4$ up to possible logarithmic corrections. Within the MV model, this limiting behavior was shown analytically in \cite{Kovner:1995ts,Gyulassy:1997vt}. 
In the saturation regime at small $k_T\lesssim Q_s^{\rm Pb}$, we find the expected $\sim 1/k_T^2$ behavior for a system with two differing saturation scales, with a flatter distribution at very low $k_T \lesssim Q_s^{\rm proton}$. This behavior was discussed in detail in \cite{Dumitru:2001ux,Blaizot:2004wv} and was previously observed numerically in \cite{Krasnitz:2000gz}. 

\begin{figure}[htb]
   \begin{center}
     \includegraphics[width=8cm]{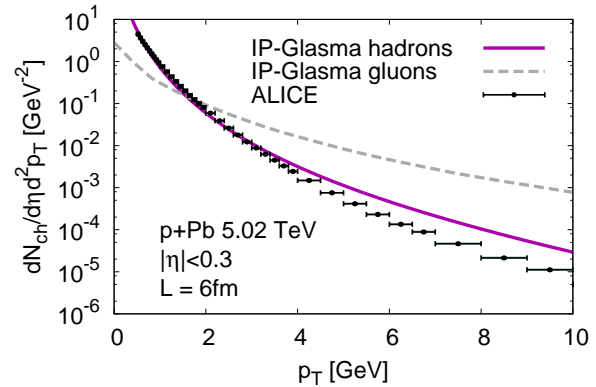}
     \caption{(Color online) Charged particle multiplicity in pPb collisions as a function of transverse momentum obtained using Eq.\,(\ref{eq:frag})
       compared to experimental data from the ALICE collaboration \cite{ALICE:2012mj}.}
     \label{fig:Npt-pPb}
   \end{center}
\end{figure}

In Fig.\,\ref{fig:Npt-pPb} we show the $p_T$ distribution of charged hadrons after fragmentation with the KKP fragmentation function, as in the p+p case.
The result is similar to the one in p+p collisions, with a good description for $p_T\lesssim 3\,{\rm GeV}$ but an overestimate in
the high $p_T$ region.

The $p_T$ integrated distribution has the functional form $(Q_s^{\rm min})^2 \ln( Q_s^{\rm max}/Q_s^{\rm min})/\alpha_S$, where $\{Q_s^{\rm min},Q_s^{\rm max}\}$ denote respectively the smaller and larger saturation scales at a given rapidity. It was shown previously to give good agreement with RHIC deuteron-gold data~\cite{Kharzeev:2002ei}. A compilation of predictions in various saturation models--all computed with the $k_T$ factorization approximation--agree with the LHC p+Pb rapidity distribution to within 20\%~ \cite{Albacete:2013ei}. 

\begin{figure}[htb]
   \begin{center}
     \includegraphics[width=8.5cm]{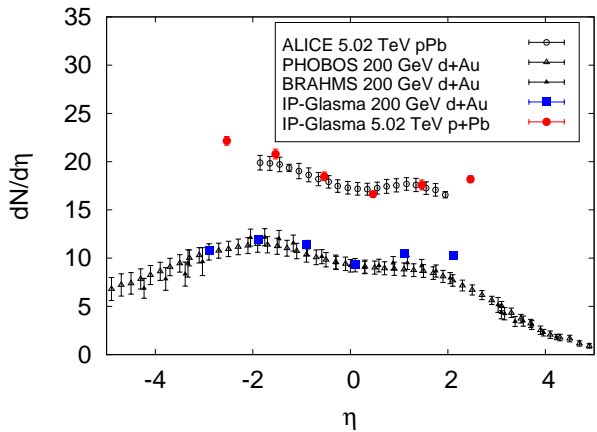}
     \caption{(Color online) Charged particle multiplicity as a function of rapidity in d+Au collisions at $200\,{\rm GeV}$ and p+Pb collisions at $5.02\,{\rm TeV}$ compared to experimental data from the PHOBOS and BRAHMS collaborations \cite{Back:2003hx,Arsene:2004cn} and the \mbox{ALICE} collaboration \cite{ALICE:2012xs}.}
     \label{fig:dNdeta-pPb}
   \end{center}
\end{figure}

In Fig.\,\ref{fig:dNdeta-pPb} we present results in the IP-Glasma framework for rapidity distributions in d+Au and p+Pb collisions.\footnote{Note our discussion of the effect of the choice of scale in the running coupling in Section \ref{sec:pp}.}
We have approximated the shift of the rapidity to the laboratory frame, in which the data is presented, by a shift of the same amount in pseudo-rapidity. In d+Au the shift is 0.11 units of rapidity in the proton going direction, in p+Pb it is 0.465 units.

It is important to note that all but one parameter are the same as in p+p collisions (thick lines in Fig.\,\ref{fig:dNdeta-pp}). The only exception is the energy dependent normalization used in the plot. We find $\mathcal{N} \approx 0.14 \ln(\sqrt{s}/1\,{\rm GeV})$, so that at a given energy the normalization constant $\mathcal{N}$ is approximately 15\% smaller than that for p+p collisions. However, this is well within the systematic uncertainties of our framework. 
The rapidity dependence is somewhat flatter than the data in d+Au collisions and slightly steeper in the higher energy p+Pb collisions. At large absolute values of $\eta$ either target or projectile are probed in the large-$x$ region that we have little theoretical control over. We thus do not expect a very good description in the very forward and backward directions.

Finally, we compare the scaled multiplicity distribution obtained within the IP-Glasma model to the preliminary scaled $N_{\rm track}$ distribution measured by the CMS collaboration \cite{CMS:2012qk,Chatrchyan:2013nka}. The data is uncorrected. We are therefore not making a comparison to correct quantity, but it is still useful to see if the shape of the distributions are similar. The results suggest reasonable agreement for up to a few times the mean multiplicity but undershoot the \emph{uncorrected} data for larger values. Nevertheless, because the agreement is reasonable in the multiplicity range in p+Pb collisions where azimuthal anisotropy moments $v_{2,3}$ become large, the gluon field fluctuations which generate these distributions (whether in the initial state or final state) are captured in the model. Our results for the multiplicity fluctuations in p+p and p+A collisions, when combined with previous results for A+A collisions~\cite{Schenke:2012hg}, therefore provide a firm basis for systematic studies of 
$v_n$ moments in small sized systems.

\begin{figure}[htb]
   \begin{center}
     \includegraphics[width=8.5cm]{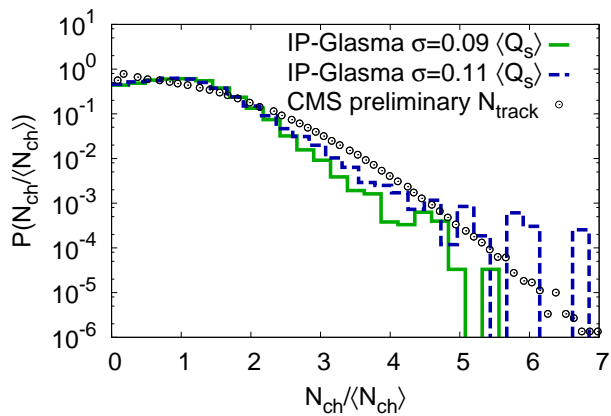}
     \caption{(Color online) Scaled distribution of the number of tracks $N_{\rm track}$ in the range $p_T>0.4\,{\rm GeV}$ and $|y|<2.4$
       for p+Pb collisions at $\sqrt{s}=5020\,{\rm GeV}$.
       The solid line is the IP-Glasma result including fluctuations of the number of gluons in the incoming nucleons, characterized by fluctuations
       of $g^2\mu^2$ at a given $Q_s^2$ the same as in $p+p$ collisions. The dashed line is the result for Gaussian fluctuations with a larger variance.}
     \label{fig:dNdy-pPB}
   \end{center}
\end{figure}

\section{Conclusions}\label{sec:conc}
In this paper, we presented several refinements to the IP-Glasma model for nucleus-nucleus collisions, which allows for a more consistent QCD motivated treatment of these collisions. The parameters in the model are clearly stated, and their variation with energy and the number of participants is explored. Further systematic studies are presented in the appendix. The sensitivity of results to model parameters is largest for peripheral nuclear collisions. 
However, this sensitivity of multiplicities to non-perturbative physics is primarily reflected in the overall normalization.
The variation of the energy and centrality dependence of multiplicities with the non-perturbative parameters of the model is relatively weak, typically less than $\sim 15\%$. 
Thus our study shows that the dominant features of the dynamics are controlled by a semi-hard scale that grows with energy and system size.
The fact that we obtain good agreement with the experimental data is therefore indicative of the key role of the saturation scale in high energy collisions.

We extended the IP-Glasma model to study the smaller sized systems produced in proton-proton and proton-nucleus collisions. 
We computed particle multiplicity distributions as a function of collision energy, rapidity, and transverse momentum, as well as the probability distribution of charged hadron multiplicities, in p+p and p+A/d+A collisions, within the IP-Glasma model. Our results show that a consistent description can be achieved over a very wide range of collision systems and energies. We find though that one has to include a further source of QCD fluctuations beyond those included in the IP-Glasma framework. While we were able to model these phenomenologically, our study points to the importance of a better theoretical understanding of these rare fluctuations, and their impact on observables such as the striking ridge correlations seen in both p+p and p+A collisions.

The IP-Glasma+\textsc{music} framework is at present the only framework that combines i) sub-nucleon scale fluctuations constrained by HERA inclusive and diffractive data, ii) Yang-Mills dynamics of non-equilibrium Glasma fields, and iii) viscous event-by-event relativistic hydrodynamics. 
A path to improving each of these elements is clear. For i) we need to solve the Balitsky-JIMWLK equations that include multi-parton correlations, 
for ii), extend the Yang-Mills treatment to 3+1-dimensions thereby incorporating essential physics of plasma instabilities and multiple scattering~\cite{Berges:2013eia,Gelis:2013rba}, and for iii), explore the temperature dependence of shear and bulk transport coefficients, among other refinements. The calculations presented here serve as an important baseline for future phenomenological studies of the wealth of data 
now available for a wide range of energies, centralities and colliding nuclei. Because the model includes both initial and final state scattering effects, such a consistent treatment, when compared to data, can help untangle their dynamics and relative importance. We see the studies in this paper as a first step in this direction.

\section*{Acknowledgments}
BPS\ and RV\ are supported under DOE Contract No. DE-AC02-98CH10886. 
This research used resources of the National Energy Research Scientific Computing Center, which is supported 
by the Office of Science of the U.S. Department of Energy under Contract No. DE-AC02-05CH11231, and additional 
computer time on the Guillimin cluster at the CLUMEQ HPC centre, a part of Compute Canada HPC facilities.

\appendix

\section{Dependence of multiplicities on model parameters}\label{sec:app1}
\begin{figure}[htb]
   \begin{center}
     \includegraphics[width=8cm]{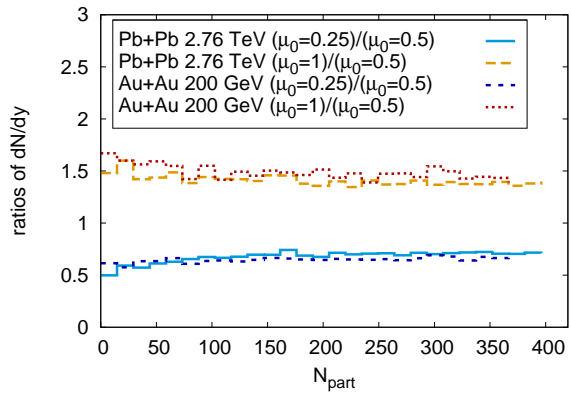}
     \caption{(Color online) Dependence on the scale $\mu_0$ below which the running coupling $\alpha_s(\tilde{\mu})$ is frozen. $m=0.1\,{\rm GeV}$ and $r_{\rm max}=1.2\,{\rm fm}$. Running with $\tilde{\mu}=0.5\,k_T$.}
     \label{fig:dNdy-mu0}
   \end{center}
\end{figure}

In this appendix, we report on further studies of the dependence of multiplicities on model parameters and parameters corresponding to lattice discretization. 
First, we study the dependence of  $\mu_0$ in Eq.\,(\ref{eq:running}). We let the coupling run as $\tilde{\mu}=0.5\,k_T$ for the purposes of this study. The results are shown in Fig.\,\ref{fig:dNdy-mu0}, where we plot the ratios of results obtained with $\mu_0=0.25\,{\rm GeV}$ and $\mu_0=1\,{\rm GeV}$ to results obtained using $\mu_0=0.5\,{\rm GeV}$. The most significant effect is an increase in multiplicity as we go to larger $\mu_0$. This occurs because, in this case, the coupling is frozen at a smaller value; note that it is the inverse of $\alpha_s(\tilde{\mu})$ that multiplies the multiplicity. The effect is slightly larger at low $N_{\rm part}$. Despite this effect on the overall normalization, one can infer from the plot that this has no effect on the energy dependence of the results. Parameter $c$ in Eq.\,(\ref{eq:running}) determines how sharply the running coupling is frozen. Varying it by factors of 2 around the standard value $c=0.2$ did not change the results. 
\begin{figure}[htb]
   \begin{center}
     \includegraphics[width=8cm]{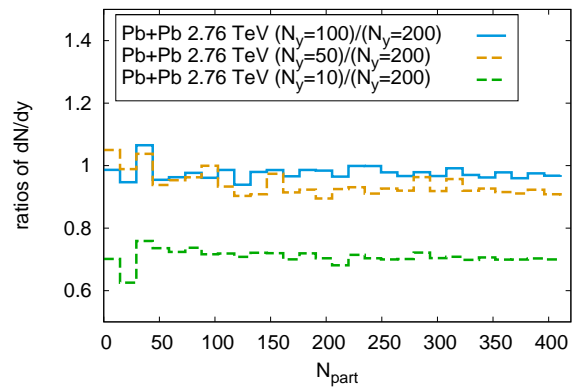}
     \caption{(Color online) Dependence on the discretization of the integral in $x^-$, given by the number of steps $N_y$.
       $m=0.1\,{\rm GeV}$ and $r_{\rm max}=1.2\,{\rm fm}$. Running with $\tilde{\mu}=0.5\,k_T$.}
     \label{fig:dNdy-Ny}
   \end{center}
\end{figure}
\begin{figure}[htb]
   \begin{center}
     \includegraphics[width=8cm]{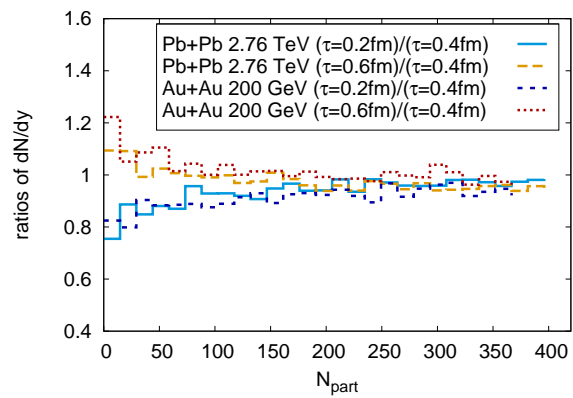}
     \caption{(Color online) Dependence on the evolution time at which the multiplicity is measured. 
       $m=0.1\,{\rm GeV}$ and $r_{\rm max}=1.2\,{\rm fm}$. Running with $\tilde{\mu}=0.5\,k_T$.}
     \label{fig:dNdy-time}
   \end{center}
\end{figure}
\begin{figure}[t]
   \begin{center}
     \includegraphics[width=8cm]{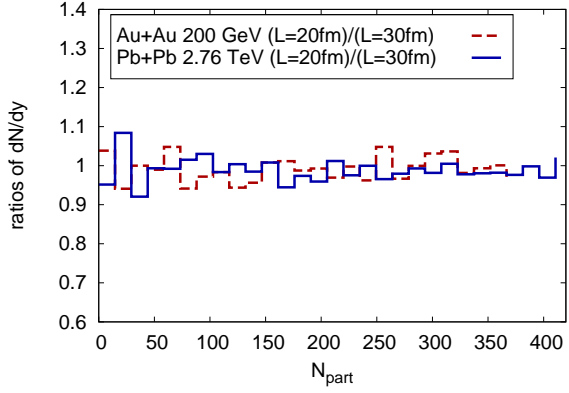}
     \caption{(Color online) Dependence on the lattice length $L$ at fixed lattice spacing $a=0.05\,{\rm fm}$.
       There is no dependence on $L$.
       $m=0.1\,{\rm GeV}$ and $r_{\rm max}=1.2\,{\rm fm}$. Running with $\tilde{\mu}=0.5\,k_T$.}
     \label{fig:dNdy-L}
   \end{center}
\end{figure}
\begin{figure}[b]
   \begin{center}
     \includegraphics[width=8cm]{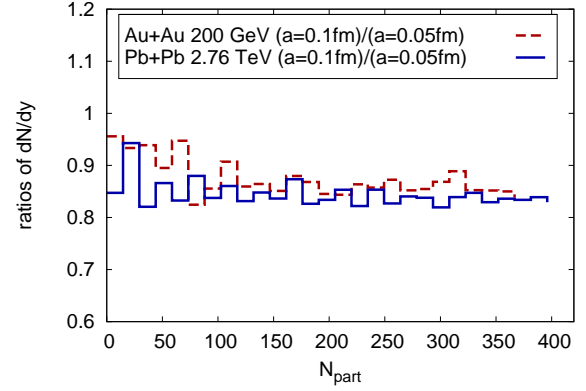}
     \caption{(Color online) Dependence on the lattice spacing $a$ for fixed $L=30\,{\rm fm}$.
       $m=0.1\,{\rm GeV}$ and $r_{\rm max}=1.2\,{\rm fm}$. Running with $\tilde{\mu}=0.5\,k_T$.}
     \label{fig:dNdy-a}
   \end{center}
\end{figure}

In Fig.\,\ref{fig:dNdy-Ny} we study the effect of $N_y$, the discretization in the $x^-$ direction in Eq.\,(\ref{eq:poe}), on the energy and centrality dependence of the multiplicity. We find that the normalization is affected for small $N_y \lesssim 50$, but for $N_y \gtrsim 100$ results have converged.
The effect is similar in Au+Au collisions at $200\,{\rm GeV}$ such that the energy dependence is only very weakly dependent on the value of $N_y$.

In Fig.\,\ref{fig:dNdy-time} we present the dependence on the time at which the multiplicity is measured. At low $N_{\rm part}$ we find a $20\%$ increase of the multiplicity from $0.2$ to $0.4\,{\rm fm}$ and then again from $0.4$ to $0.6\,{\rm fm}$. For $N_{\rm part}>200$, we see that the  results have converged after $0.2\,{\rm fm}$. Again, even though there is a centrality dependence to the time evolution, there is no energy dependence to the results. 

Finally, we show how sensitive our results are to the lattice length $L$ and the lattice spacing $a$. Fig.\,\ref{fig:dNdy-L} shows that there is virtually no dependence on the lattice length when the lattice spacing is kept constant at $a=0.05\,{\rm fm}$. This is to be expected because the infrared behavior is regulated by the mass term $m$.

Fig.\,\ref{fig:dNdy-a} shows a dependence of the overall normalization on $a$, caused by the fact that larger $a$ results in a smaller maximal transverse momentum. This clearly  affects the extraction of the 
the integrated multiplicity. 
However the shape of the integrated multiplicity as a function of $N_{\rm part}$ is almost not affected for the values of $a$ used. Neither is there an energy dependence of the ratio of multiplicities extracted at different lattice spacing.

\vspace{-0.5cm}
\bibliography{spires}

\end{document}